\pdfoutput=1
\documentclass[prl,aps,twocolumn]{revtex4}
\usepackage{bm}
\usepackage[]{graphicx}
\usepackage{amsmath}

\begin{document}

\title{Measurement of Cytoplasmic Streaming in {\it Chara Corallina}
\\ by Magnetic Resonance 
Velocimetry}

\author{Jan-Willem van de Meent,$^{1,2}$ Andy J. Sederman$^{3}$, Lynn F. Gladden$^{3}$, 
and Raymond E.  Goldstein$^{2}$}
\affiliation{$^{1}$Instituut-Lorentz/Leiden Institute of Physics, University of 
Leiden, Postbus 9506, 2300 RA, Leiden, NL}
\affiliation{$^{2}$Department of Applied Mathematics and Theoretical
Physics, University of
Cambridge, Wilberforce Road, Cambridge CB3 0WA, UK}
\affiliation{$^{1}$Department of Chemical Engineering and Biotechnology, 
University of Cambridge, Pembroke Street, Cambridge CB2 3RA, UK}

\begin{abstract}

\end{abstract}

\maketitle

{\bf In aquatic plants such as the Characean algae, the force generation 
that drives cyclosis is localized within the cytoplasm, yet produces fluid 
flows throughout the vacuole.  For this to occur the tonoplast must transmit 
hydrodynamic shear efficiently.  Here, using magnetic resonance velocimetry, 
we present the first whole-cell measurements of the cross-sectional longitudinal 
velocity field in {\it Chara corallina} and show that it is in quantitative 
agreement with a recent theoretical analysis of rotational cytoplasmic 
streaming driven by bidirectional helical forcing in the cytoplasm, with direct 
shear transmission by the tonoplast.}

\medskip


\medskip

Abbreviations: MRV, magnetic resonance velocimetry; MRI, magnetic resonance imaging;
RF, radio-frequency

\bigskip

\hrule

\bigskip

The pioneering work of Kamiya and Kuroda (Kamiya and Kuroda 1956) established that 
cytoplasmic streaming in Characean algae takes place not only in the cytoplasm, 
but also produces fluid flows that extend throughout the entire vacuole.  As 
pointed out long ago (Pickard 1972) this result implies that the vacuolar 
membrane (tonoplast) efficiently transmits shear generated in the cytoplasm 
(Houtman et al. 2007) to the vacuole (Figure 1a).  While the ability of membranes 
to flow under shear is well-known from the ``tank-treading" of erythrocytes 
(Fischer et al. 1978), it is clear from recent studies of lipid vesicles under 
extensional flows (Kantsler et al. 2007) that sheared membranes can undergo 
instabilities that lead to complex dynamics.  With the ability to image the 
dynamics of the tonoplast directly in Arabidopsis by means of a 
green-fluorescent-protein-labelled tonoplast integral channel protein 
(Cutler et al. 2000), there is fairly clear evidence that such complex dynamics 
do take place {\it in vivo}, and these may have implications for flows throughout 
the cell.  In {\it Chara corallina}, for example, the actin-myosin system 
driving streaming is localized in two opposed helical bands at the cell periphery, 
and the regions of high shear where they meet may create complex tonoplast dynamics. 

\begin{figure}[b]
\begin{center}
\includegraphics*[clip=true,width=0.8\columnwidth]{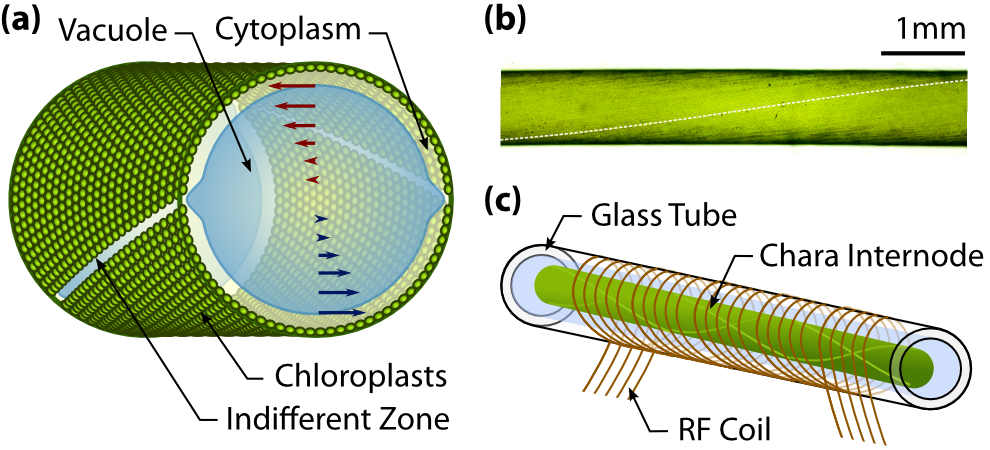}
\end{center}
\caption{\label{fig1} 
Geometry of {\it Chara corallina}. (a) Cross-section of a {\it Chara} internode. 
Flow is driven in opposing directions along two helical bands separated by 
indifferent zones visible as light lines on the surface. Velocities are generated in 
a $10$ $\mu$m thick layer of cytoplasm at the periphery, but a shear flow extends 
into the central vacuole that takes up 95\% of the volume of the cell. (b) 
Microscopy image of the sample prior to insertion into the glass tube. 
The dotted line shows a spiral of dimensionless wavelength $\lambda/R = 42$. 
(c) Schematic of sample holder enclosed in a horizontal RF coil. The length of 
the tube is $40$ mm.}
\end{figure}

\begin{figure*}[t]
\begin{center}
\includegraphics*[clip=true,width=1.7\columnwidth]{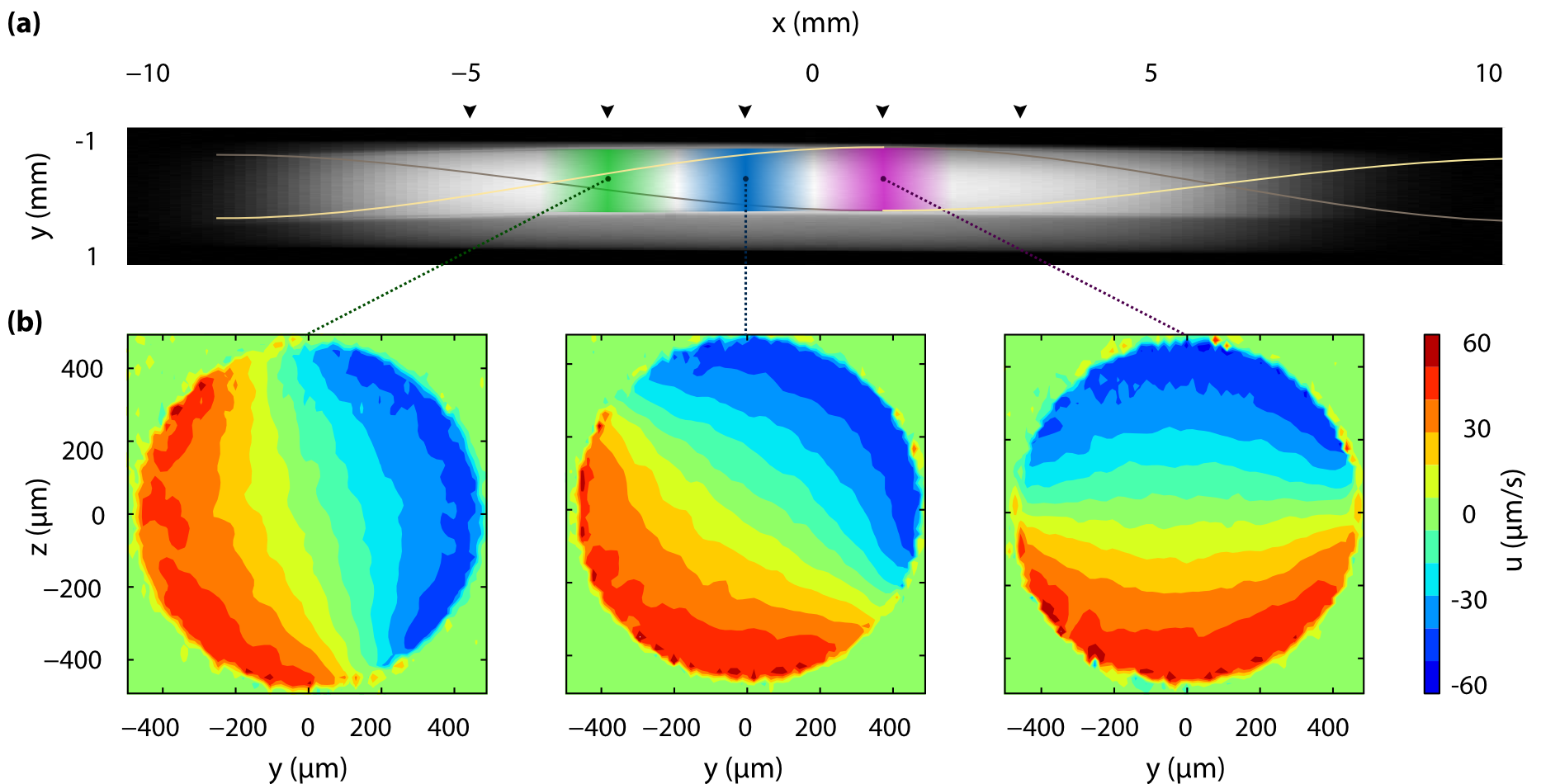}
\end{center}
\caption{\label{fig2} 
MRV measurements of the cross-sectional velocity profile of {\it Chara corallina}. 
(a) NMR image of the sample, with the measurement volumes at $x = -3$ mm , $-1$ mm, 
$1$ mm marked in green, blue and magenta. Dotted lines mark the extrapolated position 
of the indifferent zones, as determined from the orientation of the velocity profiles 
at $5$ measurement points between $x = -5$ mm and $x = 3$ mm (marked with black 
arrowheads). (b) Cross-sectional plots of the velocity. The resolutions along the $y-$ 
and $z-$axis are $16$ $\mu$m and $31$ $\mu$m respectively. Measurements were averaged 
along the $x-$axis over a Gaussian-weighted region of width $1$ mm (full-width at 
half maximum). The fitted maximum velocities of the profiles are $45.2$ $\mu$m/s, 
$46.0$ $\mu$m/s and $46.9$ $\mu$m/s.}
\end{figure*}

The possibility that streaming may impact on cellular metabolism by mixing 
the contents efficiently (Hochachka 1999, Pickard 2006) has been revisited 
recently through solutions of the coupled advection-diffusion dynamics of 
rotational cytoplasmic streaming (Goldstein et al. 2008, van de Meent et al. 2008).  
These calculations of velocities throughout the vacuole assumed that shear 
transmission by the tonoplast was complete, and showed that vacuolar flows can 
produce enhanced mixing, providing a possible mechanism for the homeostatic 
role hypothesized by Hochachka.  The first direct measurements of the 
wall-to-wall profile are those of Kamiya and Kuroda (1956), who studied rhizoid 
cells, `leaf' cells sprouting from nodes, and internodal cells, and found a constant 
velocity within the cytoplasm and a curved shear profile within the vacuole.  
Mustacich and Ware (1976) improved on these measurements by using laser-Doppler 
scattering through a chloroplast-free window obtained by exposure to an argon 
laser prior to observation. Pickard (1972) obtained velocity measurements for a 
collection of native particles in an internodal cell of {\it Chara braunii},  
showing consistency with the velocity deduced under the approximation of 
non-helical indifferent zones.

Here we report a technical advance in the study of cytoplasmic streaming by 
obtaining the fluid velocity throughout a cross-section of internodal cells 
in {\it Chara corallina} directly with magnetic resonance velocimetry (MRV).  
Our results allow for quantitative tests of recent fluid dynamical theories 
(Goldstein et al. 2008, van de Meent et al. 2008) and suggest further uses 
for magnetic resonance imaging (MRI) in the study of large-scale streaming flows.

In recent years, MRI techniques have increasingly found use in non-medical 
applications (Elkins and Alley 2007), and the advent of phase-shift 
MRV has made it possible to perform non-invasive flow measurements on 
microscopic scales (Callaghan 1994). In MRV the initial application of a 
pulsed magnetic field gradient encodes each spin with a `label' describing 
its position along the direction of the gradient. At a time $\Delta$ later 
(the ``observation time"), a reversed gradient is applied, introducing a net 
phase shift in the orientation of the nuclear spin system that is directly 
related to the distance travelled in the direction of the gradient. 
By careful selection of the magnitude of the applied field gradients and the 
observation time, velocities from $\sim 10^{-5} - 10$ m/s can be measured, 
depending on fluid properties. With sufficient time-averaging, spatial resolutions 
of $10-100$ $\mu$m can be achieved, allowing imaging of biological systems on 
scales just slightly larger than those of typical single cells (Choma et al. 2006). 
While this technique has successfully been applied at tissue level in a variety 
of plant systems (Scheenen et al. 2001, Köckenberger 2004, Windt et al. 2006), 
it is in the uniquely sized internodal cells of {\it Chara} that we can obtain 
measurements of flows internal to single cells.  

\begin{figure*}[t]
\begin{center}
\includegraphics*[clip=true,width=1.3\columnwidth]{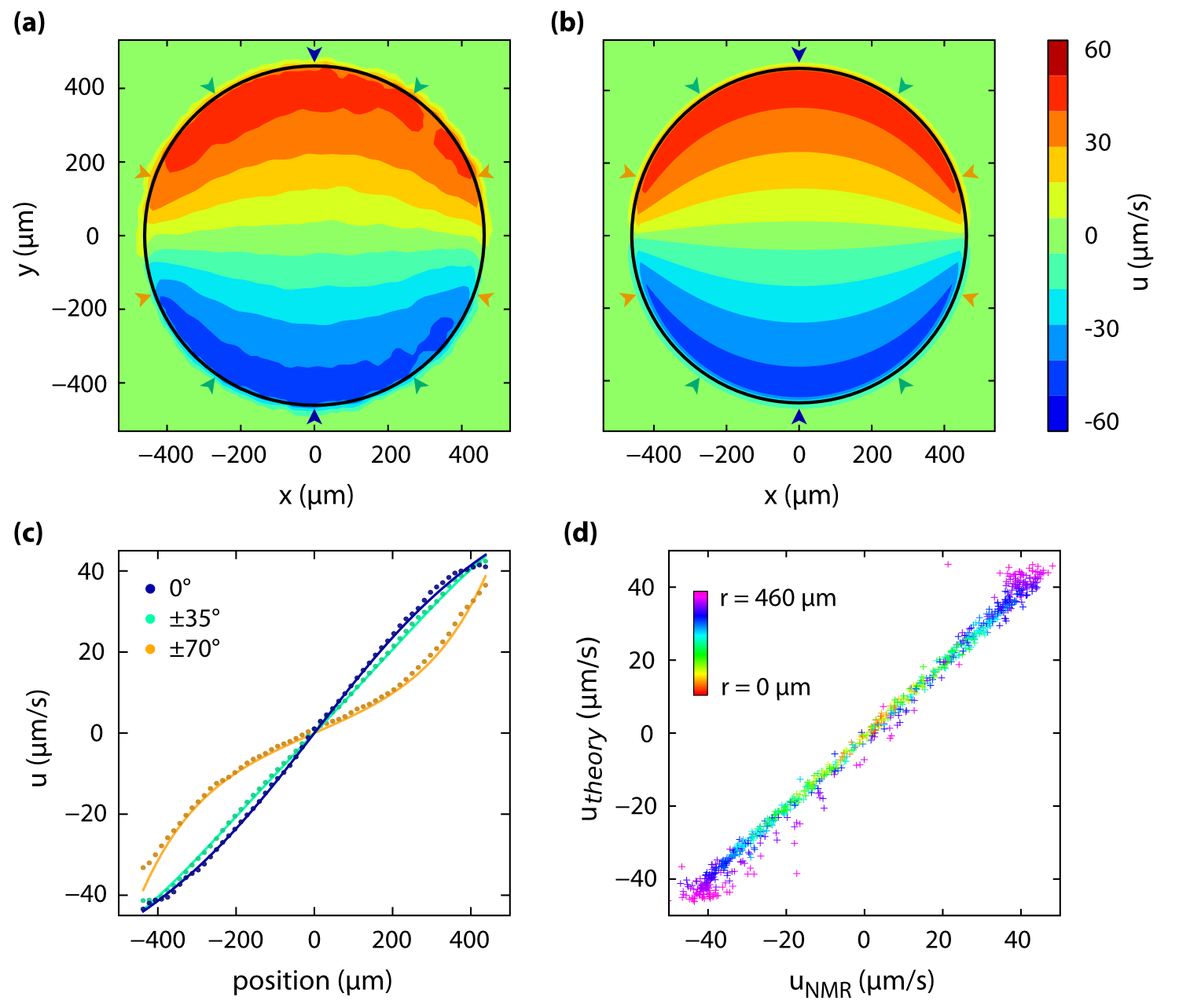}
\end{center}
\caption{\label{fig3} 
Comparison between measurements and theoretical solution. (a) Reconstructed 
profile created by aligning and averaging the data sets shown in Figure 2b. 
The black circle shows the position of the cell wall. (b) Corresponding theoretical 
profile. (c) Three wall-to-wall velocity sections at orientations of $0^{\circ}$, 
$\pm 35^{\circ}$ and $\pm 70^{\circ}$ with respect to the $z-$ axis (see arrowheads 
in panels above). For non-zero angles the data is averaged over the section 
with positive and negative orientation. Markers show the measured values, lines 
show the theoretical prediction. (d) Comparison between measurements and theoretical 
values. Data points from all three profiles are shown as markers, with the 
color of the marker denoting the radial position of the data. For clarity, 
only $1$ in $10$ points is shown.}
\end{figure*}

The Characean internode (Fig. 1a) is a single cylindrical cell with a diameter 
up to $1$ mm and a length that can exceed $10$ cm. The bulk of the volume of the 
cell is occupied by a large central vacuole, within which a $0.15$ M concentration 
of salts results in the $5$ bars of osmotic pressure that lends the cell its 
rigidity. A layer of cytoplasm $\sim 10$ $\mu$m in thickness encloses the vacuole 
at the cell periphery. Charophytes are recognised as the highest genetic 
predecessors to vascular plants (Karol et al. 2001) and in {\it Chara} 
most organelles common to higher plants are found in the cytoplasm.  
Somewhat uniquely, the millions of chloroplasts cover the cell wall, packed into 
helical rows that spiral along the inner surface (Fig. 1a-b). On the inside of 
those rows, bundled actin filaments act as tracks for myosins that drag 
structures within the cell (Kachar 1985, Kachar and Reese 1988) and thereby 
entrain cytoplasm. With streaming rates as high as $100$ $\mu$m/s, the myosin 
XI found in {\it Chara} is the fastest known (Shimmen and Yokota 2004). As a result 
of a reversed polarity of the actin filaments, the flow is organised in two 
opposing bands, producing a ``barber-pole" velocity at the cell periphery. These 
bands are separated by indifferent zones, identifiable by the absence of 
chloroplasts and visible as two light lines crossing the surface of the cell. 

To measure the cross-sectional flow inside the internodal vacuole, a sample is 
inserted into a horizontal solenoidal radio-frequency (RF) coil (Fig 1c). Flow 
profiles with a cross-sectional resolution of $16$ $\mu$m $\times 31$ $\mu$m 
are obtained over a total period of $192-256$ minutes, averaging over a volume of 
$1$ mm along the longitudinal axis of the cell (see Materials and methods).  
Figure 2a shows the location of the five averaging domains on a particular sample, 
and Fig. 2b shows the velocity profiles at three of those positions. Mean flow 
velocities are $\sim 45$ $\mu$m/s, typical of a cell of radius $R=460$ $\mu$m at 
the temperature of $298$ K inside the coil (Pickard 1974). The helical pitch 
obtained from five short measurements $2$ mm apart (Figure 2a, arrowheads), is 
$18.6^{\circ}$/mm, giving a dimensionless wavelength of $\lambda/R = 42.0$, in 
good agreement with our estimate from the light-microscopy image in Figure 1b.

Aligning and averaging the three datasets in Figure 2b we further enhance the 
signal to noise ratio of the profile. The resulting cross-section (Figure 3a) 
is in excellent agreement with the hydrodynamic solution presented in earlier 
work (Goldstein et al 2008), shown in Figure 3b. In order to account for the 
$\sim 20^{\circ}$ of helical rotation along the averaging volume, we average 
the theoretical expression (see Materials and methods) over a length of $3$ mm with 
the same Gaussian weighting as the MRV measurements. The wall-to-wall 
$1$-dimensional profiles in Figure 3c show close agreement with the theoretical 
solution. In Figure 3d, measurement points are plotted against their theoretical 
values, in the manner of earlier velocity measurements (Pickard 1972). The color 
of the points indicates the radial distance from the centre of the cell. We see 
a remarkably good correspondence throughout the bulk of the cell, with deviations 
restricted primarily to points within a pixel from the cell wall, where 
partial-volume effects become significant.

In summary, we have shown that magnetic resonance velocimetry can yield 
detailed information on the velocity distribution of cytoplasmic streaming within 
single plant cells, allowing quantitative comparison with fluid dynamical theories.  
Natural extensions to this work include studies of other streaming geometries 
found in nature and of the spread of tracers as a probe of mixing 
(Esseling-Ozdoba et al. 2008). 

\vfil
\eject

\noindent{\bf Materials and methods}

\smallskip

\noindent{\it Plant materials}

{\it Chara corallina v. australis} was obtained from the Botanic Garden of the 
University of Cambridge, courtesy of J. Banfield, and grown in a non-axenic 
culture, rooted in non-fertilized soil in a 100 liter tank filled with Artificial Pond 
Water (1 mM NaCl, 0.4 mM KCl, 0.1 mM CaCl$_2$). The tank was kept at 
room temperature and illuminated with a bench lamp on a 16/8 hour day-night cycle. 
During illumination, the light intensity at the top of the tank was $\sim 250$ lux. 
Samples of suitable size were placed in a Petri dish under a microscope to 
verify healthy streaming.

In preparation for measurements, the sample was inserted into a $3.0$ mm outer 
diameter, $1.6$ mm inner diameter glass capillary $40$ mm in length, which was 
pre-filled with Forsberg medium (Forsberg 1965). The capillary tube was closed 
with PDMS plugs and a small volume of silicone grease to ensure a good seal. 

\smallskip

\noindent{\it Magnetic resonance velocimetry}

Velocimetry experiments were performed on a Bruker Spectrospin DMX 200, 
4.7 T magnet with a $20$ mm long solenoid coil of diameter $3$ mm. $^1$H images 
were acquired at $199.7$ MHz. Spatial resolution was achieved using $3$-axis 
shielded gradient coils providing a maximum gradient strength of $49$ G cm$^{-1}$ 
in each direction. Transport is measured over the observation time, $\Delta$, and 
since the RMS displacement increases as $\Delta^{0.5}$ due to diffusion and 
as $\Delta^1$ due to convection, for short $\Delta$ diffusive (incoherent) 
displacements can dominate over convective (coherent) ones, particularly in 
slowly convecting systems. Therefore, to weight the measurement towards the 
convective field a stimulated echo sequence was used to enable a large observation 
time for motion encoding.  Further spatial imaging gradients were applied after 
the motion encoding to minimize diffusive attenuation. Hard $90^{\circ}$ RF pulses 
were used except for the final pulse which was a Gaussian-shaped selective 
$90^{\circ}$ RF pulse $512$ $\mu$s in duration. Experimental parameters used for the 
velocity images were: observation time, $\Delta = 500$ ms; velocity gradient 
duration, $d = 1.62$ ms; gradient increment, $g_{\rm inc} = 10$ G cm$^{-1}$; 
number of velocity gradient increments, $2$; recycle time, $TR=1.9$ s; 
number of scans = $16$; field-of-view = $2$ mm $\times$ $2$ mm; pixel array size 
$N_{\rm read} \times N_{\rm phase} = 128 \times 64$; in-plane spatial resolution 
= $16$ $\mu$m $\times$ $31$ $\mu$m; slice thickness = $1$ mm; 
measurement duration = $64$ min. 

For each data set in Figure 2b, $4$ measurements were taken between $3$ and $7$ hours 
apart and subsequently averaged. A Gaussian smoothing with $31$ $\mu$m FWHM was 
selectively applied to the streaming region whilst avoiding blurring at the cell 
periphery. The streaming velocity in the individual measurements was not observed 
to vary significantly over the $15$ hour period of acquisition.  Throughout the 
measurements the temperature was maintained at $298$ K by a regulated 
airflow system.

\smallskip

\noindent{\it Hydrodynamic solution}

The hydrodynamic solution was obtained as a mode expansion of the equations for 
Stokes flow (Goldstein et al. 2008). The system of equations assumes an infinitely 
long helical cell, allowing simplification of the problem to a $3$-dimensional flow 
with symmetry, where the velocity depends only on the radial coordinate $r$ and a 
helical angle $\varphi = \theta - (2\pi/\lambda) x$. Here, $\theta$ is the polar 
angle in the $yz$-plane and $\lambda$ is the wavelength of the helical bands.

The solution of the flow problem can be expressed as a sum over a series of 
modes $f_i(r) \sin((2i+1) \varphi)$, with the radial modes $f_i(r)$ given in 
terms of a sum of Bessel functions. In the results presented here, $64$ modes 
were used for the expansion. The velocity at the boundary was taken 
piece-wise constant on the bands, with a smooth crossover at the indifferent zones 
given by $\tanh(\varphi/\epsilon)$ and $\tanh((\pi-\varphi)/\epsilon)$ respectively. 
The cross-over width used was $\epsilon = 11$ $\mu$m.

To allow better comparison with the MRV  measurements, a series of profiles was 
averaged along the $x$-axis over a length of $3$ mm, using a Gaussian weighting with 
a spread of $1$ mm full-width at half maximum. 

\medskip

\noindent{\bf Funding}

This work was supported by the Engineering and Physical Sciences Research Council 
[DTA to J.W.vdM., EP/F047991/1 to A.J.S. and L.F.G.], the Biotechnology and Biological 
Sciences Research Council [BB/F021844/1 to R.E.G.], Leiden University 
[to J.W.vdM.], the Leverhulme Trust and the Schlumberger Chair Fund [to R.E.G.].

\medskip

\noindent{\bf Acknowledgments}

We are grateful to M. Polin, T.J. Pedley, C. Picard, and I. Tuval for numerous discussions.

\bigskip

\noindent{\bf References}

Callaghan, P.T. (1994) {\it Principles of Nuclear Magnetic Resonance Spectroscopy}. 
Oxford University Press, Oxford.

Choma, M.A., Ellerbee, A.K., Yazdanfar, S. and Izatt, J.A. (2006) Doppler flow 
imaging of cytoplasmic streaming using spectral domain phase microscopy.  
{\it J. Biomed. Optics} 11:024014.

Cutler, S.R., Ehrhardt, D.W., Griffitts, J.S. and Somerville, C.R. (2000) 
Random GFP::cDNA fusions enable visualization of subcellular structures in cells of 
{\it Arabidopsis} at a high frequency. {\it Proc. Natl. Acad. Sci. USA} 97: 3718-3723.

Elkins, C.J. and Alley, M.T. (2007) Magnetic resonance velocimetry: applications of 
magnetic resonance imaging in the measurement of fluid motion. {\it Exp. Fluids} 43:823-858.

Esseling-Ozdoba, A., Houtman, C., van Lammeren, A.A.M., Eiser, E. and Emons, A.M.C. 
(2008) Hydrodynamic flow in the cytoplasm of plant cells. {\it J. Microscopy} 231:274-283.

Fischer, T.M., Stohr-Lissen, M. And Schmid-Schonbein, H. (1978) The red cell as a fluid 
droplet: tank tread-like motion of the human erythrocyte membrane in shear flow. 
{\it Science} 202:894-896.

Forsberg, C. (1965) Nutritional Studies of Chara in Axenic Cultures. {\it Physiologia 
Plantarum} 18:275-290.

Goldstein, R.E., Tuval, I. and van de Meent, J.W. (2008) Microfluidics of cytoplasmic 
streaming and its implications for intracellular transport. {\it Proc. Natl. Acad. 
Sci. USA} 105:3663-3667.

Hochachka, P.W. (1999) The metabolic implications of intracellular circulation. 
{\it Proc. Natl. Acad. Sci. USA} 96:12233-12239.

Houtman, D., Pagonabarraga, I. Lowe, C.P., Esseling-Ozboda, Emons, A.M.C., and Eiser, E. 
(2007) Hydrodynamic flow caused by active transport along cytoskeletal elements. 
{\it Europhys. Lett.} 78:18001.

Kachar, B. (1985) Direct visualization of organelle movement along actin-filaments dissociated 
from Characean algae. {\it Science} 227:1355-1357.

Kachar, B. and Reese, T.S. (1988) The mechanism of cytoplasmic streaming in Characean 
algal cells -- sliding of endoplasmic-reticulum along actin-filaments. {\it J. Cell. Biol.} 
106:1545-1552.

Kamiya, N. and Kuroda, K. (1956) Velocity distribution of the protoplasmic streaming in 
{\it Nitella} cells. {\it Bot. Mag. Tokyo} 69:544-554.

Kantsler, V., Segre, E. and Steinberg, V. (2007). Vesicle dynamics in time-dependent 
elongational flow: wrinkling instability. {\it Phys. Rev. Lett.} 99:178102.

Karol, K.G. et al. (2001) The closest living relatives of land plants. {\it Science} 
294:2351-2353.

K{\"o}ckenberger, W., De Panfilis, C., Santoro, D., Dahiya, P. and Rawsthorne, S. 
(2004), High resolution NMR microscopy of plants and fungi. {\it J. Microscopy} 214:182-189.

Mustacich, R.V. and Ware, B.R. (1976) A study of protoplasmic streaming in {\it Nitella} 
by laser Doppler spectroscopy. {\it Biophys. J.} 16:373-388.

Pickard, W.F. (1972) Further observations on cytoplasmic streaming in {\it Chara braunii}. 
{\it Can. J. Bot.} 50:703-711.

Pickard, W.F. (2003) The role of cytoplasmic streaming in symplastic transport. 
{\it Plant, Cell. Environ.} 26:1-15.

Pickard, W.F. (2006) Absorption by a moving spherical organelle in a heterogeneous 
cytoplasm: implications for the role of trafficking in a symplast. {\it J. Theor. Biol.} 
240:288-301.

Scheenen, T.W.J., Vergeldt, F.J., Windt, C.W., de Jager, P.S. and Van As, H. (2001) 
Microscopic imaging of slow flow and diffusion: a pulsed field gradient stimulated echo 
sequence combined with turbo spin echo imaging. {\it J. Mag. Res.} 151:94-100.

Shimmen, T. and Yokota, E. (2004) Cytoplasmic streaming in plants. 
{\it Curr. Op. Cell Biol.} 16:68-72.

van de Meent, J.W., Tuval, I. and Goldstein, R.E. (2008) Nature's microfluidic 
transporter: rotational cytoplasmic streaming at high P{\'e}clet numbers. 
{\it Phys. Rev. Lett.} 101:178102.

Windt, C.W., Vergeldt, F.J., de Jager, P.A. and van As, H. (2006). MRI of long-distance 
water transport: a comparison of the phloem and xylem flow characteristics and dynamics in 
poplar, castor bean, tomato and tobacco. {\it Plant Cell \& Envir.} 29:1715-1729.

\end{document}